\documentclass[fleqn,10pt]{wlscirep}
\usepackage[utf8]{inputenc}
\usepackage[T1]{fontenc}

\usepackage{lineno}
\usepackage{booktabs}
\usepackage{multirow}
\usepackage{longtable}
\usepackage{xr-hyper}
\usepackage[charter,cal]{mathdesign}

\makeatletter
\newcommand*{\addFileDependency}[1]{
  \typeout{(#1)}
  \@addtofilelist{#1}
  \IfFileExists{#1}{}{\typeout{No file #1.}}
}
\makeatother

\newcommand*{\myexternaldocument}[1]{%
    \externaldocument{#1}%
    \addFileDependency{#1.tex}%
    \addFileDependency{#1.aux}%
}
\myexternaldocument{si}

\title{
Extreme heat reduces and reshapes urban mobility
}

\author[a,*]{Andrew Renninger}
\author[b]{Carmen Cabrera}

\affil[a]{Centre for Advanced Spatial Analysis, University College London, London, UK}
\affil[b]{Geographic Data Science Lab, University of Liverpool}

\affil[*]{Corresponding author: Andrew Renninger (E-mail: andrew.renninger.12@ucl.ac.uk)}

\begin{abstract}
Extreme heat is a problem in European countries and cities, with rising temperatures affecting ageing populations. Research on mobility during extreme heat remains limited to small samples and isolated contexts, leaving significant gaps in our understanding how entire populations adjust their day-to-day activities and how these adaptations vary across social groups. Here we use data from passive and active mobile network connections covering 13 million individuals in Spain (27\% of the population) to examine extreme heat's impact on mobility at scale. We stratify by age, gender, economic class, and activity. Our findings show mobility falls by as much as 10\% on hot days generally and 20\% on hot afternoons specifically, when temperatures peak. Further differences emerge on hot days. Older adults cut travel to work and other activities, while those earning less are less able to avoid work; social mixing declines and spatial structure changes as activity falls in city centres. These disruptions have implications for urban economies, as curbed activity and interaction—both planned and unplanned—threaten the dynamism of cities as hubs of social and economic exchange. 
\end{abstract}
\begin{document}

\flushbottom
\maketitle

\section*{Introduction}
Extreme heat poses a serious threat to lives, livelihoods, and the economy \cite{zhao2024global, masselot2023excess}. Rising temperatures have been linked to increased hospital admissions \cite{zhang2015impact, schaffer2012emergency} and increased mortality rates \cite{vicedo2021burden, kaiser2007effect}. Extreme heat reduces productivity in both manufacturing \cite{somanathan2021impact} and agriculture \cite{schlenker2009nonlinear}, and slows economic growth \cite{dell2014we, colacito2019temperature}. These challenges have been compounded by the growing intensity and duration of heat waves over the past century \cite{perkins2012increasing, perkins2020increasing}. Here we develop our understanding of how daily travel responds to extreme heat by linking mobility data, stratified by a rich set of socioeconomic attributes, with climatic conditions. Our findings provide the first estimates of how different populations adapt to heat, revealing constraints and trade-offs while documenting broad disruptions in urban life.

The effects of extreme heat are uneven across populations, as some groups are more vulnerable or exposed than others \cite{zhao2024global, kenny2010heat}. For example, the elderly are disproportionately affected by extreme temperatures \cite{xu2014impact, oudin2011heat} due to increased vulnerability, often associated with chronic conditions such as diabetes \cite{WHO2023Diabetes, WHOObesityChallenge}, which heighten their risk during heat events \cite{zanobetti2012summer}. Workers in physical labor \cite{romanello20212021}, such as construction and agriculture \cite{cruz2024global}, face significant risks due to prolonged exposure to extreme temperatures. Socioeconomic status determines adaptive capacity, as wealthier households turn on air conditioning at lower temperatures than poorer households \cite{cong2022unveiling}. The widespread adoption of air conditioning has reduced heat-related mortality, highlighting the importance of wealth in adapting to climate stress \cite{barreca2016adapting, heutel2021adaptation}.

Individuals respond to hot weather based on their level of vulnerability and exposure to high temperatures. For example, time-use studies suggest that individuals reduce outdoor activities and shift towards indoor spaces during heat waves \cite{batur2024understanding, graff2014temperature}, yet such shifts may be constrained by income, occupation, or urban design \cite{graff2014temperature}. Mobility decisions represent important means of adaptation \cite{li2024urban, gu2024socio, bocker2013impact, basu2024hot}. Cities see more cyclists and pedestrians on warm days than on hot and humid days \cite{bocker2019weather, bocker2013impact}, while more employees miss work those who do work are less productive \cite{zander2015heat}. Social and economic incentives may influence decisions, with the risk of heat stroke heightened for the military and during athletic competitions \cite{periard2022exertional}. Further, shocks to infrastructure may either encourage or discourage adaptation, with railways, roadways, and energy grids experiencing greater strain during heat waves \cite{ferranti2016heat}. 

Critical questions remain about how heat shapes human mobility across different populations \cite{bocker2013impact}, with digital trace data now enabling us to study these behavioural dynamics at scale \cite{hatchett2021mobility}. These data allow us to monitor how populations respond to extreme weather events, from fires to floods \cite{gonzalez2024harnessing, li2022spatiotemporal}, and recent work reveals differences in how socioeconomic groups responded to extreme heat during the pandemic \cite{ly2023exploring}. Because most work is limited in scope and scale, we lack a strong understanding of how heat disrupts and restructures urban life. Cities thrive on interactions \cite{storper2004buzz, arzaghi2008networking, berkes2021geography, duranton2004micro}, and systematic reductions in these interactions could suppress the benefits of agglomeration. 


The European context provides a valuable setting with historically temperate climate but a sharp rise in extreme heat events \cite{yin2019mapping}. This combination of historically moderate temperatures and increasing heat exposure creates a natural laboratory for studying adaptation. Compared to many other regions, Europeans show vulnerability to heat at lower temperatures \cite{tobias2021geographical}, and—contra evidence for adaptation cite{hess2023public}—recent heat waves have been equally fatal as those in prior decades \cite{ballester2023heat, gallo2024heat}. Europe’s ageing population, with growing rates of chronic conditions like diabetes \cite{WHO2023Diabetes, WHOObesityChallenge}, exacerbates the risk from heat, as both age \cite{xu2014impact, oudin2011heat} and chronic illness \cite{zanobetti2012summer} are associated with greater risk from extreme heat. The continent’s low adoption of air conditioning compared to regions like the United States \cite{IEA2018} also creates risks and forces populations to rely on behavioural adaptations to cope with rising temperatures. 

Spain, where we focus here, exemplifies these European dynamics while facing intensifying challenges. Projections for the next 50 years suggest that Southern Europe, including Spain, will experience a combination of rising temperatures, increased drought frequency, and ageing infrastructure \cite{wu2024temperature}. Heat waves in this region are expected to become not only more intense but also more spatially expansive \cite{lorenzo2021heatwave}, affecting larger portions of the peninsula during any given event. These challenges highlight the urgent need for studies that can inform adaptation strategies tailored to European cities and populations.

Here we examine the effect of extreme heat on daily mobility patterns in Spain, thereby shedding light on the economic and human impacts of high temperatures in cities. We combine large-scale mobility data collected in 2021 and 2022 with high-resolution estimates of thermal comfort. By focusing on Spain, we explore how extreme heat alters activity patterns in a European context, considering factors such as the type of activity as well as demographic and socioeconomic disparities. We find a clear distinction between routine and non-routine activities, with the latter disrupted more than the former. Commensurate with greater risk from heat, the oldest populations adjust more than the youngest; less affluent groups change mobility less than the wealthy. Critically, we document how extreme heat systematically reorganizes urban network structures and reduces socioeconomic mixing, threatening the core mechanism of cities as "social reactors" \cite{bettencourt2013origins} where the density and diversity of interaction generates scaling advantages in productivity and innovation. Our research offers new insights into how populations manage the challenges of extreme heat, trading off relative risks and needs, and informs adaptation strategies for European cities.

\section*{Results}
To understand the effect of heat on activity we start by linking data on thermal comfort with data on daily mobility. We then apply several modeling techniques to detect statistically significant changes in mobility behaviour across various population groups.

We use data provided by the Spanish Ministry for Transport \cite{mitma}, which contains records for the movements of $\sim13$ million individuals, or $\sim27\%$ of the population. The data represent flows within and between 3,999 districts, including mainland Spain and the Balearic Islands, and are stratified according a variety of important characteristics, including a broad classification of the activity and the trip distance, as well as the age, gender and economic class of the person making the trip. Trips are logged from both active events like texts and calls as well as passive events in the form of probes from the network operator, allowing high temporal and spatial resolution. Because they come from network operators rather than applications, these data have comparably less bias than data from aggregators of GPS location data and they are validated and balanced with surveys and administrative statistics to ensure quality and reliability (see Methods for more detail). Fig. \ref{fig1}\textbf{A} shows the networks for each month, demonstrating its strong coverage in both urban and rural areas, and we show time series and validation in Supplementary Figs. S1 and S2. The data that we leverage here allow us to decompose travel according to demographic attributes without resorting to imputation. Because the data are aggregated into districts, we cannot see the precise activity, but each activity is given a category: home, work/study, and frequently- or infrequently-visited places. According to the Ministry for Transport, an activity is ``frequent'' if a given person visits that place more than once in two weeks, and it is ``infrequent'' if not. 

We measure daily variations in experienced heat using ERA5-HEAT data \cite{di2021era5}, which gives the Universal Thermal Climate Index (UTCI). This metric reflects perceived ambient conditions by incorporating temperature, humidity, wind, and solar radiation into a standardised formula. Focusing specifically on the impact of extreme heat, our analysis is restricted to the summer months, defined as May through September, for the years 2021 and 2022. Fig. \ref{fig1}\textbf{B} show the mean UTCI across months, with the strongest temperatures in July and August. Spain experienced significant heat waves in both periods, with 2022 seeing waves from June through August, while 2021 had milder conditions with record highs in August.

\subsection*{Heat drives a sharp decline in travel for discretionary activities, with minimal impact on work or study mobility}

\begin{figure*}[bt!]
\centering
\includegraphics[width=1\textwidth]{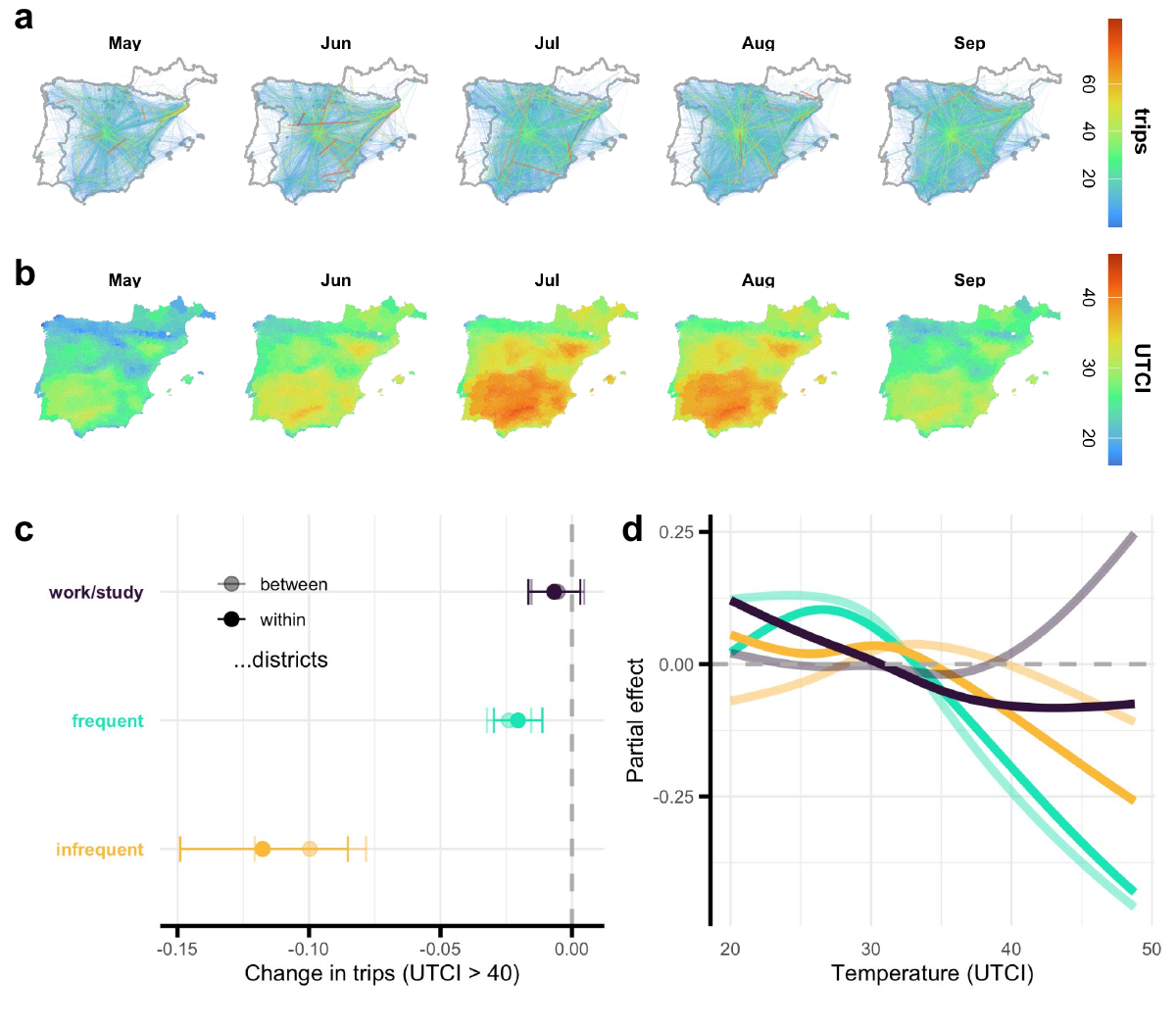}
\caption{\textbf{The effect of extreme heat on activity.} \textbf{A} Mobility networks in Spain across months in 2022 and 2023; because the data represent 30\% of the population, the networks have good coverage, including rural and urban areas, although the network is dominated by cities like Madrid in the middle of the country. \textbf{B} Mean temperatures in the same period, with heat peaking in July and August. \textbf{C} Estimates from a two-way fixed effects model, controlling for district and date, on the effect of a UTCI above $40^{\circ}$C, showing that frequent and infrequent activities fall while trips to work or school hold steady. \textbf{D} When we examine effect of temperature along the continuum rather than in binary, we see that for both frequent and infrequent activities, and there are warm temperatures that increase activity and hot temperatures that decrease them.}
\label{fig1}
\end{figure*}

We use a two-way fixed effects model (TWFE) to estimate the causal effect of temperature on activity by exploiting variation in temperature while controlling for both time-invariant attributes and spatially-uniform shocks through district and date fixed effects. Shown in Fig. \ref{fig1}\textbf{C} (and reported in Supplementary Table T1), infrequent trips fall around 10\% and trips to work or school see little change, with frequent trips falling almost 3\%. Placebo tests, where we shuffle UTCI either by date or district, in Supplementary Fig. S5, show no effect for permuted UTCI on activity, indicating that the results are not spurious. 

To identify the form of the relationship between temperature and activity, we turn to a generalised additive model (GAM), which fits smooth functions to capture nonlinear relationships in data and in doing so extract effects across different temperatures. Our specification is similar to above, but we fit a cubic spline by day-of-year to model the seasonality, and include day-of-week and holiday terms to account for variations across days. Showing this continuous relationship between heat and activity in Fig. \ref{fig1}\textbf{D}, we see that higher temperatures result in lower activity. Although this model disagrees with the TWFE in which activities are most responsive to heat, the GAM shows that there is a temperature at which activity in both categories peaks. (In Supplementary Fig. S4, we show these patterns are consistent when we limit our sample to large cities or small cities, and when we hold out 2022 or 2023.) This suggests that as climate patterns shift over time, some seasons and places will see visits and trips increase as warmer weather generates activity while others will see them decrease as hotter weather destroys activity. 

In both models, work and study are not responsive to heat. Further, infrequent trips within a district—those that are more likely to be traversed on foot—fall more than those between districts. Next we leverage the rich demographic and geographic attributes to examine how these effects vary across different populations and contexts. 

\subsection*{Extreme heat affects mobility most for the elderly and poor, with no significant gender differences}

\begin{figure*}[bt!]
\centering
\includegraphics[width=1\textwidth]{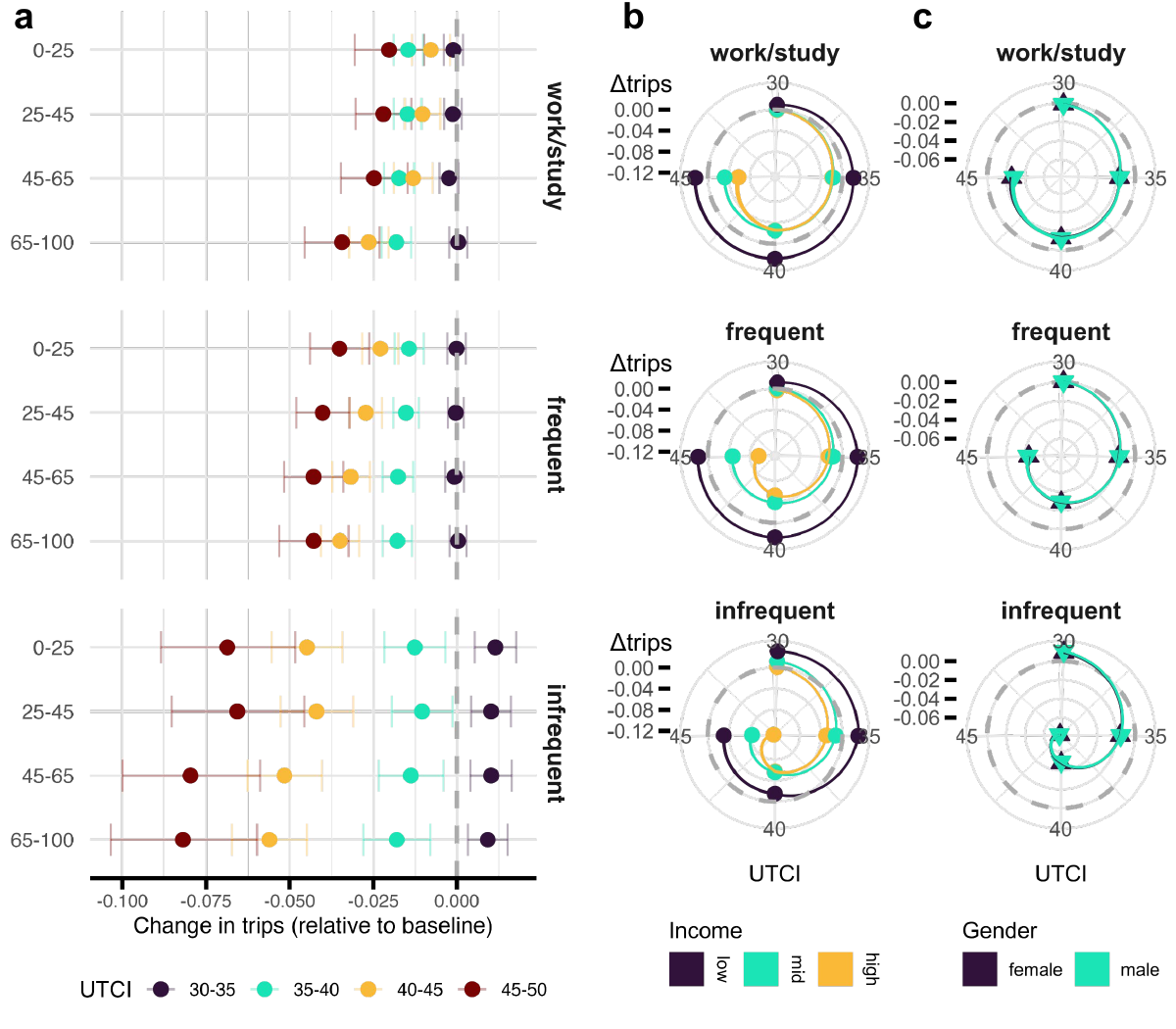}
\caption{\textbf{The effect of extreme heat on different groups.} \textbf{A} Disaggregating by age, we see that as temperatures increases and the regularity of the activity decreases, trips decline, but more for the oldest than the youngest people. We also note that warm-not-hot temperatures generate infrequent activities, people may expand their repertoire of activities during clement weather. \textbf{B} Change in effect size as UTCI increases for different income brackets, showing that the poor less sensitive to high temperatures than the wealthy, which may be tied to work-from-home, although all groups reduce infrequent activities. This suggests that discretionary mobility changes but obligations like work do not. \textbf{C} We see no differences by gender, with activity falling identically as UTCI rises for both males and females despite aggregate differences in mobility between the genders.}
\label{fig2}
\end{figure*}

In addition to activities, we disaggregate according to various demographic attributes, including age, gender and income. Our preferred specification is the TWFE which, under certain assumptions (see Methods section), allows us to estimate a causal effect of extreme heat on mobility. Here we use temperature bands rather than a binary indicator of extreme heat. Our estimates are both statistically and practically significant, showing reductions across all classes of activity. The patterns we see in this section vary systematically between activities and monotonically across temperatures, lending confidence to the relationship we see across point estimates: higher temperatures mean lower mobility. 

In Fig. \ref{fig2}\textbf{A} we see a gradient, with higher temperatures corresponding to stronger declines in activity across all age groups. Looking at how different ages respond to extreme heat, our results are clear: mobility for the young is the least affected by high temperatures and the impact becomes larger as age increases. For the oldest group, a given day with heat index above $45^{\circ}$C corresponds to an 8\% decline in infrequent activity, a 4\% decline in frequent activity, and even a 3\% change in work or study. Those in middle age visit work 2.5\% less on the hottest days. Because we are using data from 2022 and 2023, we note that these effects do not necessarily mean a reduction in work, because the foregone travel could be to work from home. Yet this change in behaviour would still have implications for cities as Spain warms over time: if the elderly are \emph{missing} work, there is a direct economic cost to the workers, but if many are simply working from home, then the economic burden falls on the shops and restaurants that rely on business from commuters.

This is consistent with the fact that heat poses a greater risk to older populations than it does to younger ones \cite{oudin2011heat}. In relative terms, the oldest are most affected by extreme heat but because they constitute a larger and more active population, the greatest decline in absolute terms comes from the middle-aged population, while younger populations are least impacted in both absolute and relative terms.  

There are many plausible channels by which income and heat could interact but here we propose two: the wealthy might be more capable of coping with extreme heat, via air conditioned homes and cars, and thus remain unaffected; the poor might be less able to afford missing work. White collar jobs allow at least some remote work and many blue collar jobs do not. Shown in Fig. \ref{fig2}\textbf{B}, our results indicate that the poor cannot afford, or are not able, to miss work. Individuals from households in the lowest income bracket are unaffected by high temperatures while those from households in the highest income bracket reduce travel across all classes of activity.

Supporting the hypothesis that work compels the least affluent group to stay active, we see that this group still curbs infrequent activities while holding steady trips for work or study as well as for frequent activities even when the heat index surpasses $45^{\circ}$C. These frequent activities could be attached to daily or weekly routines like lunch breaks, or taking children to school and therefore co-occur with work. For the wealthiest group, all three classes of activity fall at that level of discomfort—by as much as 10\% and 15\% for frequent and infrequent activities respectively.

With large differences in labor force participation between men and women \cite{OECD2022Employment}, and differences in both unpaid work and care work between men and women \cite{OECD2016Care}, we might expect routines to vary enough to see variation in responses to heat. Yet once we stratify on the type of activity in Fig. \ref{fig2}\textbf{C}, we observe no gender differences in mobility. (We also note in Supplementary Fig. S6 that there are no statistically significant differences between the trip counts in the network when we stratify on gender.) 

\subsection*{Larger drops in the afternoon and on short trips that may involve active travel}

\begin{figure*}[bt!]
\centering
\includegraphics[width=1\textwidth]{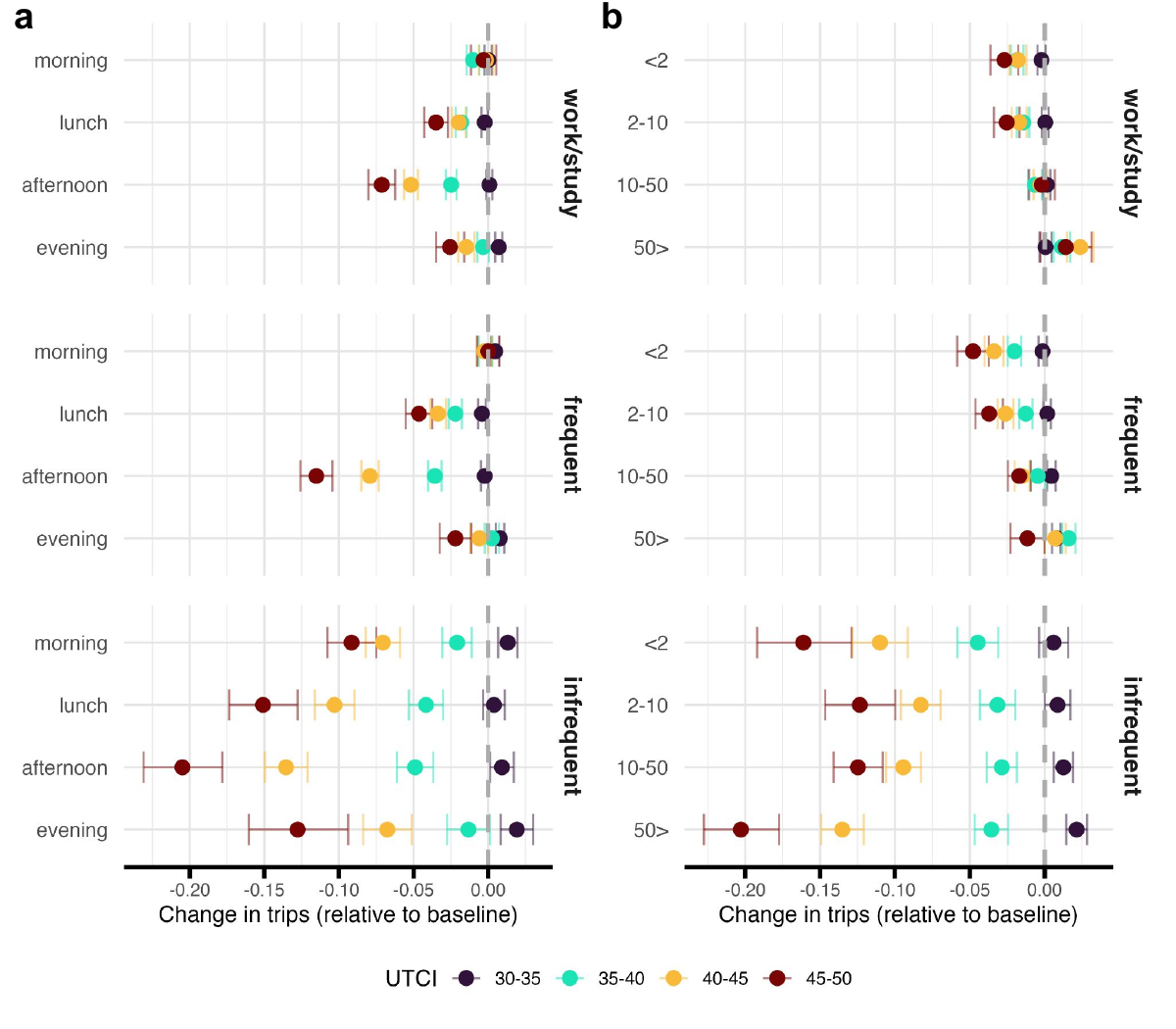}
\caption{\textbf{Time and distance.} \textbf{A} We see patterns consistent with our earlier findings, wherein trips to frequently-visited locations fall more than trips to work or school, and trips to infrequently-visited locations fall most, but we also see that this effect is pronounced in the afternoon when temperatures are crest. We do not see evidence of substitution to the morning or evening before or after temperatures peak, as those trips still decline on hot days. \textbf{B} Generally, longer trips are most resilient to extreme heat and shorter trips, and cars may play a role in this difference as shorter trips are more likely to taken on foot, but infrequent trips see strong changes at both long and short extremes.}
\label{fig3}
\end{figure*}

Because heat is variable throughout the day, starting off cooler in the morning, heating up in the afternoon and cooling later in the evening, we test whether or not people respond this progression. Fig. \ref{fig3}\textbf{A} shows that visits to all classes of activity fall more in the afternoon on hot days than they do in the morning, by as much as 20\% for infrequent activities on the hottest days of the year. Yet even frequent activities, which may be coupled with work or study, fall by more than 10\%. Taken together, this also lends credibility to our earlier estimates because it shows that mobility responds not just to hot days but the hottest part of the day, which would be less likely if we were observing a spurious effect. It also suggests that even when people go to work on a hot day, they might do fewer other activities during the those afternoons.   

In Fig. \ref{fig3}\textbf{B}, we explore effects across different journey lengths. Our data do not allow us to interrogate why people might be avoiding certain kinds of travel, but in our models we see that the largest reduction in activity that comes from trips that span less than 2 km, which are less likely to involve a car. This agrees with literature showing that cycling and walking are most impacted by hot days \cite{bocker2013impact}. Again we see tight coupling between mobility for work or study and for frequent activities, suggesting that certain activities might go hand-in-hand with a routine that includes both professional obligations and personal needs. Although long trips are generally the least affected by high temperatures, they experience the largest declines when they involve infrequent activities, specifically when UTCI exceeds $40^{\circ}$C. 

While we see evidence that long trips for infrequent activities increase when temperatures are mild and decrease when they are hot, we find no evidence that people are moving from hot to cool with long trips. In Supplementary Table T2, we modify a standard gravity model, which estimates flows between districts based on distance and population, to include the temperature gradient between origin and destination; this gradient shows no effect on flows. 

\subsection*{Reduced social mixing as temperature rises}
\begin{figure*}[bt!]
\centering
\includegraphics[width=1\textwidth]{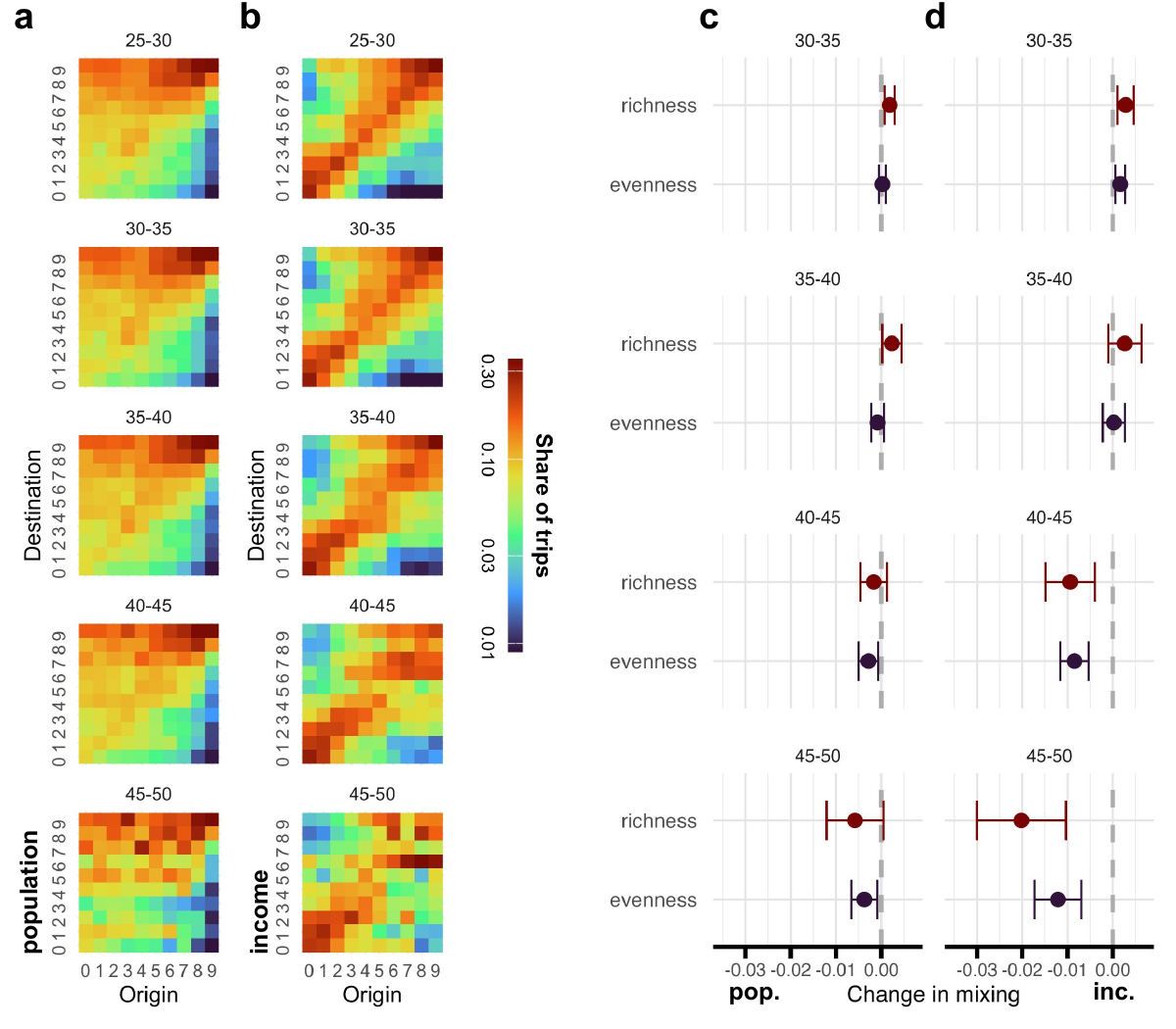}
\caption{\textbf{Mobility patterns and heat.} \textbf{A} Changes in flows between districts ordered by population decile, showing that at normal temperatures most flows go from low population to high population areas, or stay within high population districts, but this relationship breaks down during the hottest temperatures and more trips flow between what are suburban densities. \textbf{B} Changes in flows between districts ordered by median income decile, showing that there is a subtle but consistent bias wherein people from middle income districts tend to visit upper income districts, and this relationship also fades at higher temperatures—although trips continue to flow between lower income districts. \textbf{C} The results of a more controlled test, using a TWFE to predict the change in richness and evenness of visitors to a destination according to origin population, and \textbf{D} according to origin income; we see a stronger decline income diversity and the same gradient we see with total activity, with stronger temperatures generating stronger effects.} 
\label{fig4}
\end{figure*}

We also document significant changes to the structure of the mobility network as temperature changes, which may have implications for how urban areas function and how social groups mix. Fig. \ref{fig4}\textbf{A} and \ref{fig4}\textbf{B} show how trips flow to and from districts with different populations and different incomes (in deciles), respectively. Generally, trips flow from less populous to more populous areas (urban bias), and from lower income to higher income (wealth bias). These patterns are marked by the higher values in the upper triangles of the matrices in Fig. \ref{fig4}\textbf{A} and \ref{fig4}\textbf{B}. We stratify on UTCI to show how these mixing profiles vary under different climatic conditions, and plot results by row according to different UTCI bands. When it becomes hotter, these twin biases attenuate. More trips occur within middle quantiles, and in particular few trips flow from middle income to high income. Agreeing with our earlier results showing limited change amongst the poorest, flows within and between low income districts holds constant. (Note that although districts are large units, with $\sim8$ thousand residents, the ratio of \emph{between}-district to \emph{within}-district flows is 3:1, so while many needs are met within each district there is substantial potential for mixing.)

Building on these descriptive results, we introduce another TWFE design to relate mixing with temperature, using metrics borrowed from ecology \cite{pyron2010characterizing}: \emph{richness}, defined as the raw number of income or population groups who visit, and \emph{evenness}, measured using the Shannon entropy of visitor distribution across these population or income groups. The new metrics, richness and evenness, are used as the dependent variables in our TWFE model. Looking at mixing between low and high population districts in Fig. \ref{fig4}\textbf{C}, we see less of a change than what is visible in the matrices, although there is a slight reduction in mixing between rural, exurban, suburban, and urban classes at the highest extremes. Looking at mixing by economic class in Fig. \ref{fig4}\textbf{D}, however, richness falls by $\sim2\%$ and evenness falls by $\sim1.2\%$ on very hot days. Taken together, this suggests that much of the change in mobility that we see in the matrices is attributable to seasonal variations, yet it also shows a direct effect from heat on mixing. As we move from lower to higher temperatures, we see a consistent progression; we also see evidence for mixing on mild days, fitting with earlier indications that some temperatures are conducive to new activities. Taken together, these findings suggest that extreme heat not only reduces overall mobility but systematically alters the network structures that determine core-periphery interaction and socioeconomic mixing.

\begin{figure*}[bt!]
\centering
\includegraphics[width=1\textwidth]{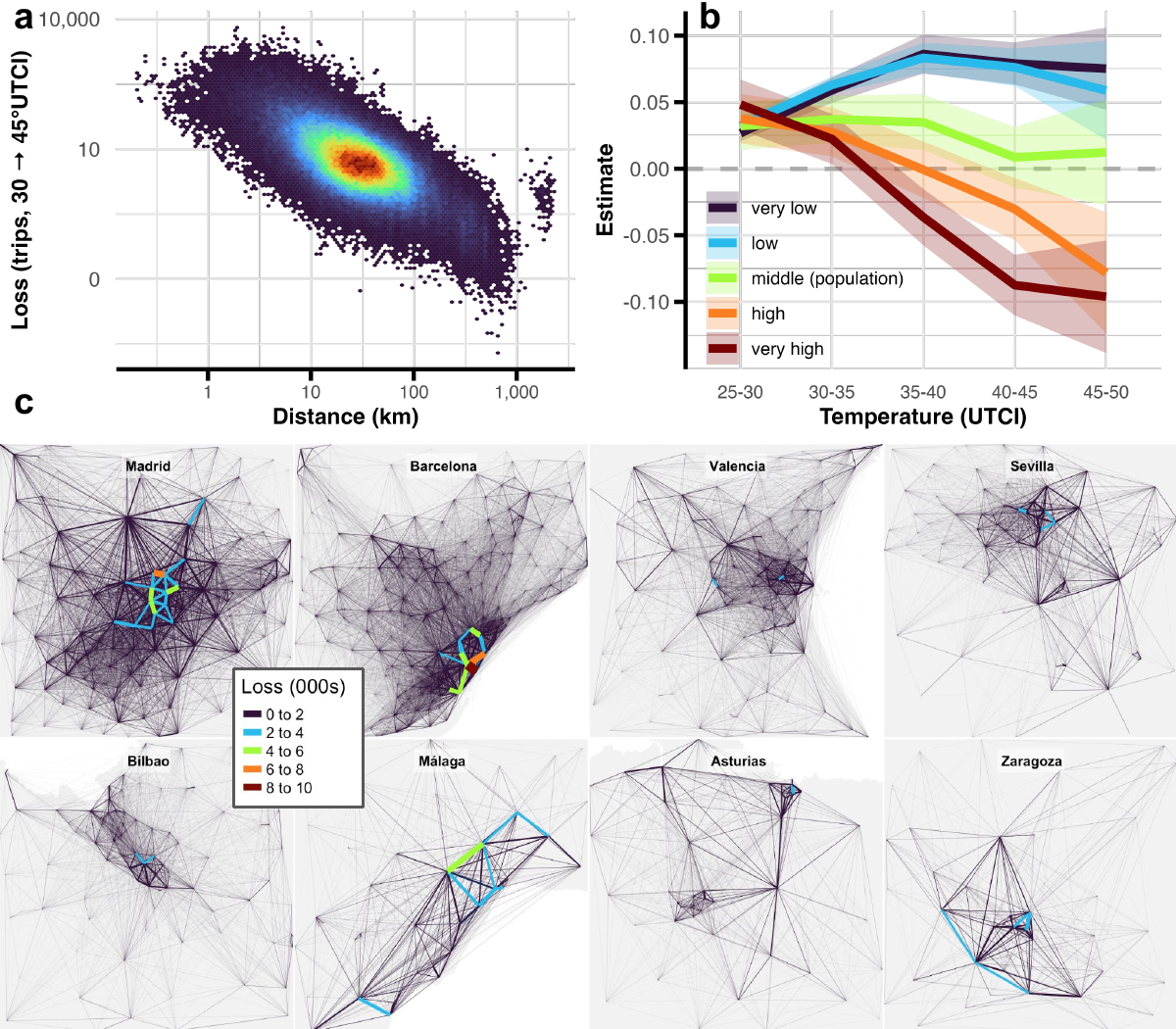}
\caption{\textbf{Urban impacts.} \textbf{A} We build a network of trips between districts and show the relationship between the loss of trips between districts and the distance between them, when temperatures rise from $35 \rightarrow 40^{\circ}$C, according to a gravity model with temperature, finding a clear negative slope. \textbf{B} In order to model how different parts the rural-urban gradient respond to heat, we add population quantiles to that gravity model, and see that more populous areas see larger reductions in trips. \textbf{C} Building out the networks from this gravity model, we can see that edges in the core see the largest declines in flows as temperatures rise from $35 \rightarrow 40^{\circ}$C, while edges in the periphery are preserved.}
\label{fig5}
\end{figure*}

We model changes to urban structure formally using a gravity model that considers population at origin and destination, along with distance between districts and the temperature on the day. The gravity coefficients, shown in Supplementary Table T3, confirm the earlier models but gives us the ability to observe changes on the network. In Fig. \ref{fig5}\textbf{A} and \textbf{B}, we show the components of that model: shorter edges see flows decline more on hot days that longer edges, and flows between populous areas fall most. Rural areas see mobility taper off from the optimum—at $\sim35^{\circ}$C but the net effect relative to cold weather is still positive. We see the consequences of this in Fig. \ref{fig5}C, in that losses on hot days tend to be concentrated in the urban core rather than the periphery. Generally, in this network analysis we see that social life in cities will be disrupted if extreme heat worsens without adaptation, with less mixing and less activity in city centres—a development which could threaten the economic and social advantages of cities.  

We use modeled estimates of Earth's future climate, derived from CMIP6 \cite{thrasher2022nasa}, which represents the state-of-the-art in climate projections (see Methods for further detail). The strategy that we employ here is simple: we switch out the UTCI for a given district and day in 2023 for the UTCI on that day in 2073, 50 years later. Holding all else constant allows us to predict activity using our fitted models. The simplicity of this approach introduces limitations, which we discuss now, but it does allow us to understand the implications of our findings in possible future with little adaptation.  

\subsection*{Cities stand to lose the most}

\begin{figure*}[bt!]
\centering
\includegraphics[width=1\textwidth]{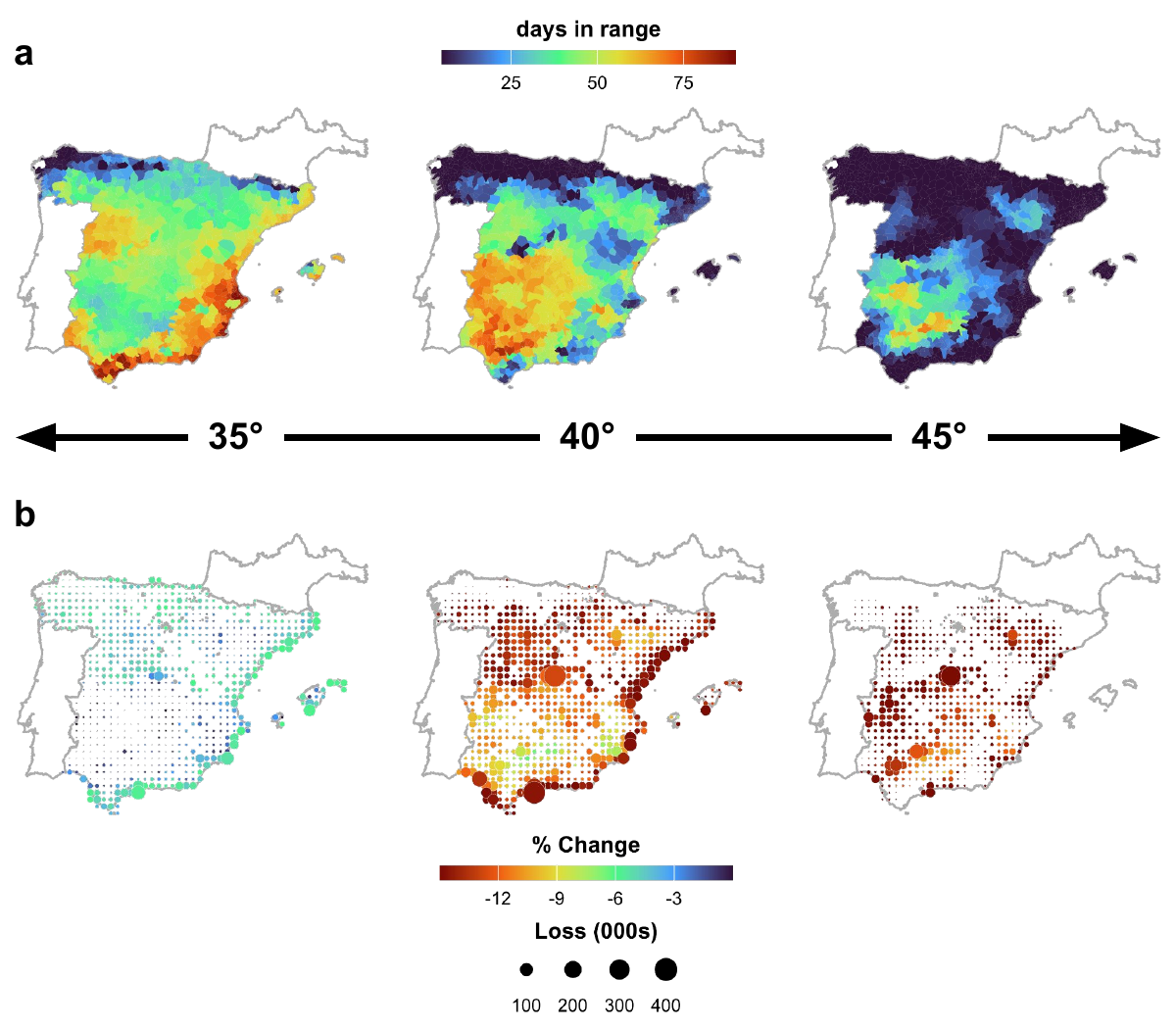}
\caption{\textbf{Projecting in the future.} \textbf{A} Prevalence of temperature exceeding a given value. Maps show that Spain's South will see 75 days between $40^{\circ}$C and $45^{\circ}$C as well as dozens of days above $45^{\circ}$C; days between $35^{\circ}$C and $40^{\circ}$C will also be more common in the North. \textbf{B} Consequences of those temperatures in both relative and absolute terms per district, showing that according to our model the largest effects at lower temperatures will be in areas that do not experience high temperatures now. Cities in the South like Sevilla and Malaga will experienced the highest temperatures, however, they do not see the strongest drop in relative terms; big cities like Madrid and Barcelona will see the largest change in absolute terms.}
\label{fig6}
\end{figure*}

Our estimates assume that temperature and nothing else changes going forward. Although the ``Lucas Critique'' \cite{lucas1976econometric} tells us to be wary of making projections when humans can adapt to changing circumstances \cite{kahn2021adapting}, Spain is ageing and thus the demographic issues that we highlight above could be exacerbated. The exercise is also informative because activity is fundamentally different to mortality and morbidity, where we do see evidence for adaptation in recent years \cite{heutel2021adaptation, barreca2016adapting}: travel within and between cities will likely require, for the foreseeable future, contact with the ambient air. Coping with heat may reduce mobility more, not less, as we attempt to reduce its worst effects on health. For example, it may become more common for employees to work from home on hot days, as telework has changed the demand for face-to-face interaction \cite{barrero2023evolution}. For this reason, we believe our estimates to have important implications for the future economy of cities and towns, including, for example, urban business that depend on office workers or rural areas that depend on tourists \cite{yabe2024behaviour}.  

Our model predicts a 3.5\% reduction in frequent activities during summer months, May to September, and a 4.7\% reduction in infrequent activities. Travel to work or school is projected to fall by 2.2\% during that same period. Yet over the full year, all activities will decline by just 1\%, as warmer weather during what were once cold periods increases mobility while hotter weather during already hot seasons decreases it.  

Next we explore how this could vary across cities and regions. Fig. \ref{fig6}\textbf{A} shows the number of days spent in a given UTCI range, from $35-40^{\circ}$C to $40-45^{\circ}$C to $45-50^{\circ}$C. All of these temperatures correspond to reductions in activity according to our models, and they will be common in the South of the country. In particular, parts of Andalusia will see temperatures exceed $40^{\circ}$C for more than 75 days each year. The North of Spain will experience more days that exceed $35^{\circ}$C, but will only have a few days a year at the extremes that the South will endure. 

Fig. \ref{fig6}\textbf{B} shows that the effects of these changes may change mobility if adaptations do not address rising temperatures. Our estimates here show the change on a given day at a given temperature, compared to the baseline. Effect sizes are larger at lower temperatures in the North, where society is not as well acclimated to heat. Because extremes that will occur in the South will not be prevalent there, we concentrate on temperatures between $35^{\circ}$C and $40^{\circ}$C; days in this range would still reduce activity by as much as 6\% on days when they occur. While the South is largely unaffected by these temperatures, it see strong changes to activity when temperatures exceed $40^{\circ}$C; when this occurs, activity in the South will fall by 6-12\% while activity in the remainder of the country could fall be 15\%.

\section*{Discussion}
We find that mobility responds to heat in ways that are consistent with expectations rooted in the literature on extreme heat and health. First, daily movement patterns for the elderly are most affected by extreme heat; second, afternoons see the greatest decline in mobility levels across population groups, as temperature—and thus risk—crests. Commensurate with heightened risk, we find that those older than 65 are more likely to reduce activity in response to high temperatures, and that their reductions are greater at higher temperatures. Compatible with the daily temperature dynamics, we see that on hot days activity falls most during the early and late afternoon, and least in the mornings and evenings. Again, our results show that as temperature increases, so does the effect size in our model estimations. These changes redound to important changes to spatial structure and social mixing in cities.    

Yet we document opposing effects from extreme heat amongst another vulnerable group. While the oldest dramatically reduce discretionary activities and skip travel to work, possibly to either work from home or miss work, the poorest do not miss work and reduce activities less. This means that the group most at risk, because age magnifies the threat of heat, is responding according to that risk, but it also suggests that the poor are least able to compensate for extreme heat by foregoing work and travel. This reveals important economic constraints that may influence mortality and morbidity.

Lending confidence to our estimates and findings here is the consistency with which our models behave: higher temperatures correspond with larger effects in all of our specifications. Although these findings appear intuitive, our study is the first to document these changes accounting for trip and individual attributes; in doing so, it demonstrates the adaptive nature of human mobility in the presence of extreme heat: populations respond to high temperatures by changing routines and avoiding certain activities. Without more granular data, we cannot shed light on what all of these activities are specifically, but we do see patterns in the analysis we are able to make. 

With an ageing population and a warming world, our results suggest that policies to adapt to extreme heat will be important for keeping Spain and possibly other European countries active and productive in the coming decades. Yet many existing strategies to mitigate the worst health effects of extreme heat involve air conditioning \cite{barreca2016adapting}, including cooling shelters \cite{vcity2024cooling}, which is difficult to apply to activity. There are still strategies to mitigate extreme heat between buildings, like greening \cite{jay2021reducing}, and certain modes of travel can also be air conditioned, but a broad drive to move people to air conditioning could change the social fabric in cities, and we find preliminary evidence of how this might occur here. The short trips that are more likely to be walked are also more likely to be avoided in extreme heat. On the hottest days, many also avoid travel to work, so businesses that depend on commuters for foot traffic might suffer. The Bohemian neighbourhoods at the hearts of cities lose the most activity. The very adaptations that make heat survivable might erode the subtle interactions that make cities engines of innovation \cite{arzaghi2008networking, atkin2022returns} and culture \cite{glaeser2011triumph}. Our findings suggest, while air conditioning and work-from-home may reduce the threat of extreme heat, cooling interventions across neighbourhoods and cities—like greening and shading, or changing paving and building materials to reduce solar absorption—may preserve the interactions that define cities.

Although our results are consistent, intuitive and suggestive, the mechanisms underlying the behavioural changes we see in the data during extreme heat are not clear. For example, we cannot prove a link between short trips and active travel, and although we know that frequent trips appear linked to work/study, which agrees with research on trip chaining \cite{miyauchi2021economics}, we cannot determine which classes of amenities are most affected when people avoid travel to work. Our study is thus limited, providing strong evidence for adaptation while much of the details are left to speculation. 

\section*{Methods}
\subsection*{Data}
We use daily origin-destination data provided by the Spanish Ministry for Transport \cite{mitma}, which integrates anonymised mobile phone records with demographic, land use, and transport network information to produce a mobility data product. This dataset captures trips over 500 meters within Spain and infers key travel characteristics, including origin and destination points, travel modes, and trip purposes—either work/study, frequent or infrequent locations for that device. The Ministry leverages state-of-the-art algorithms to transform raw mobile network data into structured and scalable matrices, offering high-resolution insights into mobility patterns across spatial and temporal scales.

While many studies using GPS location data are only able to impute demographic attributes using administrative statistics \cite{moro2021mobility, de2024people}, often aggregated to large areal units that make the estimates crude, the data are stratified to allow for interrogation variations across demographic groups and trip purposes \cite{mitma2024metodologico}. Activities are classified based on the Ministry of Transport's recurrent mobility analysis, which tracks origin-destination pairs over 2-week periods. Destinations visited more than once in this window are classified as frequent activities, while those visited only once are considered sporadic activities. Balance is achieved using official statistics from the national statistics agency to account for differences in age, income, and regional population distributions. This ensures that the dataset is representative of the broader population, minimizing biases associated with the uneven distribution of mobile phone users. The integration of demographic and geographic information also allows for the segmentation of mobility patterns by municipality, province, and other spatial units, providing a flexible foundation for granular analysis. 

The Ministry for Transport employs rigorous quality controls to ensure the reliability of these data \cite{mitma2024controles}. Anomalies in travel patterns are monitored through automated systems, which compare data against historical trends and predefined thresholds. Possible errors, such as geolocation inaccuracies or missing records, are flagged and investigated to ensure data integrity. The data are also validated with independent sources, such as FAMILITUR survey data, to confirm the consistency of observed trends with government statistics. Additionally, logical consistency checks, such as evaluating the symmetry of origin-destination flows, are conducted to ensure that the data align with expected behaviours. These efforts, combined with transparent methodological documentation, make this dataset an important resource for understanding mobility in Spain.

We link these data on mobility with an index of thermal comfort from ERA5-HEAT climate reanalysis data \cite{di2021era5}, provided by the Copernicus program. Universal Thermal Climate Index (UTCI) combines temperature, wind, radiation and humidity to measure not just how hot it is but how it feels—for example, if humidity limits evaporative cooling through perspiration. We use zonal statistics to compute the mean UTCI at 16:00 for each district on each day. 

\subsection*{Models}
We employ twin modeling strategies to understand the relationship between heat and mobility, the first to measure the causal effect and the second to estimate the functional form. Both assume the number of trips $T$ terminating in district $i$ at time $t$ follow a Poisson distribution such that $T_{it} \sim \text{Poisson}(\mu_{it})$. Our first approach uses a two-way fixed effects model (TWFE):

\begin{align*}
\log(\mu_{it}) &= \beta(\text{UTCI}_{it} \times \text{activity}) + \alpha_i + \gamma_t
\end{align*}

\noindent
where $\mu_{it}$ represents the expected number of trips, $\alpha_i$ represents district fixed effects controlling for characteristics of the \emph{place}, while $\gamma_t$ captures date fixed effects accounting for patterns common across districts at a given \emph{time}. UTCI is either binary ($>40{\circ}$C) or binned ($5^{\circ}$C intervals from $20-50^{\circ}$C). The interaction with activity type allows us to estimate differential temperature responses across activities. This specification leverages within-district variation in temperature after accounting for common temporal shocks, providing causal estimates under the assumption that temperature variation is as-good-as-random after controlling for location and time fixed effects. The district controls spatial confounds and the date controls are important to adjust for temporal patterns, as Spain sees activity change considerably in August as many people make holidays during this month. We cluster standard errors at the province level to account for spatial correlation in the error terms.

A placebo test looks for confounding variables by shuffling temperature and examining whether our model detects an effect that should not exist. Our data allow us to test both spatial and temporal confounding, and we do so by permuting the temperature either within a district, so that each district on each day is treated with temperatures from another day of the year, or within a date, so that each district on each day is treated with temperatures from a another district in a different part of Spain.

Our TWFE model enables us to make causal inferences if certain conditions are met, but there are threats to the model. A key assumption is that mobility patterns would have evolved similarly across districts in the absence of temperature shocks (parallel trends). This is plausible in our setting because temperature variation is as-good-as-random after conditioning on location and time fixed effects, and districts cannot select into heat ``treatment''. A more serious threat to identification would be if extreme heat causes people to substitute between districts, violating the stable unit treatment value assumption (SUTVA). We test for such spillovers using a gravity model of bilateral flows between districts, explained and shown in Table T2 in the SI. After controlling for origin and destination populations and distance, we find no evidence that temperature gradients between districts drive mobility patterns, suggesting SUTVA violations are unlikely to bias our estimates.

To explore and model the potential curvilinear relationship between heat and activity, we complement the TWFE analysis with a Generalised Additive Model (GAM):

\begin{align*}
\log(\mu_{it}) &= f(\text{UTCI}_{it}) \times \text{activity}_i + \beta_1\text{popularity}_i + \beta_2\text{province}_i + f(\text{DoY}_t) + \text{DoW}_t + \text{holiday}_t
\end{align*}

\noindent
where $f(\cdot)$ represents a cubic regression spline with 4 knots, $f(\cdot)$ is a cubic spline for the day-of-year, capturing seasonality and drifts in the data, and we control for mean visitation (popularity) and geographic variation (province). We also add day-of-week and holiday fixed effects because, for example, weekends and holidays might have different levels of activity and this allows the intercept to vary on those days. While the TWFE isolates the causal effect, the GAM reveals the functional form of behavioural responses to temperature variation through its flexible smooth functions. The GAM's strength lies in its ability to detect and convey nonlinear relationships without imposing \emph{a priori} assumptions about the functional form, allowing us to identify potential threshold effects and complex response patterns in human mobility. 

We also employ a spatial interaction model to understand how heat specifically impacts flow dynamics between geographic areas. The gravity model, widely used in mobility studies, provides a framework for estimating the volume of flows between pairs of locations based on their attributes and the distance between them. In our application, we include temperature bands to examine how extreme heat modifies these spatial interactions. Formally, our model specification is:

\begin{align*}
\log(\mathrm{E}[T_{ijt}]) = \beta_1\mathrm{UTCI}_{jt} + \beta_2(\mathrm{UTCI}_{jt} \times \mathrm{Pop}^q_i) + \beta_3(\mathrm{UTCI}_{jt} \times \mathrm{Pop}^q_j) + \beta_4\log(d_{ij}) + \alpha_i + \gamma_j + \delta_t
\end{align*}

\noindent
where $T_{ijt}$ represents the number of trips from origin $i$ to destination $j$ on date $t$, $\mathrm{UTCI}_{jt}$ denotes the temperature band at the destination (binned into 5°C intervals from 20-50°C), $\mathrm{Pop}^q_i$ and $\mathrm{Pop}^q_j$ are population quintiles for origin and destination districts respectively, and $d_{ij}$ represents the distance between districts. We include origin fixed effects $\alpha_i$, destination fixed effects $\gamma_j$, and date fixed effects $\delta_t$ to control for time-invariant characteristics of districts and temporal patterns. The interaction terms between temperature bands and population quintiles are particularly valuable as they allow the effect of heat to vary across the urban-rural gradient, creating different slopes for different settlement types while the fixed effects adjust the intercepts, enabling us to identify how extreme temperatures differentially impact flows between urban cores, suburban areas, and rural peripheries. This specification allows us to isolate how temperature variations modify the traditional distance decay relationship while accounting for population size—a key factor in determining trip attractiveness. We implement this as a fixed-effects Poisson model, appropriate for count data, with two-way clustering of standard errors to account for spatial correlation. This hierarchical approach enables us to decompose how extreme heat differentially affects areas along the urban-rural gradient while maintaining statistical rigor through our fixed-effects structure.

\subsection*{Projections}
In order to understand possible futures according to our models, we use climate scenarios derived from state-of-the-art ``general circulation" models \cite{thrasher2022nasa}, which are used by the Intergovernmental Panel on Climate Change to make projections about Earth's future climate. (For context, Supplementary Fig. S7 shows projections for major cities.) We extract data on temperature, humidity, radiation, and wind to compute UTCI manually for the year 2050 under the $2-4.5^{\circ}$C of warming, using the simulation from the Centre National de Recherches Météorologiques in France, because it shows high accuracy against observed data \cite{craigmile2023comparing} and of comparably accurate models it performed best in Spain when we checked it against our data for 2022 and 2023. 

Our strategy for estimating mobility in the future is simple yet crude: we replace the UTCI for a given day-of-year in 2023 with the UTCI for that same day 50 years later in 2073, predicting with the new temperatures and all else equal. We compare predicted values using 2023 temperatures to predicted values using 2073 data to ensure that we are comparing like with like, modeled estimates in both cases, rather than using observed values in one and predicted values in another. Because the GAM allows us to produce estimates across the full range of temperatures, we make our predictions using this model rather than the TWFE. 


\bibliography{heat}

\section*{Acknowledgements}
The authors would like to thank Mattia Mazzoli and \emph{Complexity72h} for introducing us to the data, and Elsa Arcaute and Neave O'Clery for helping refine the paper.

\section*{Author contributions statement}
\textbf{A.R.} Conceptualization, methodology, investigation, writing, reviewing, editing; \textbf{C.C.A.} Conceptualization, methodology, investigation, writing, reviewing, editing. 

\subsection*{Data and code availability}
Mobility data are available from the Ministry for Transport. ERA5 climate recordings and CMIP6 climate projections are accessible via Google Earth Engine. All code is available upon request. 
\section*{Competing interests}
The authors declare no conflict of interest.

\end{document}


\flushbottom
\maketitle

\tableofcontents

\clearpage


\section{Data validation}

\begin{figure*}[h!]
\centering
\includegraphics[width=1\textwidth]{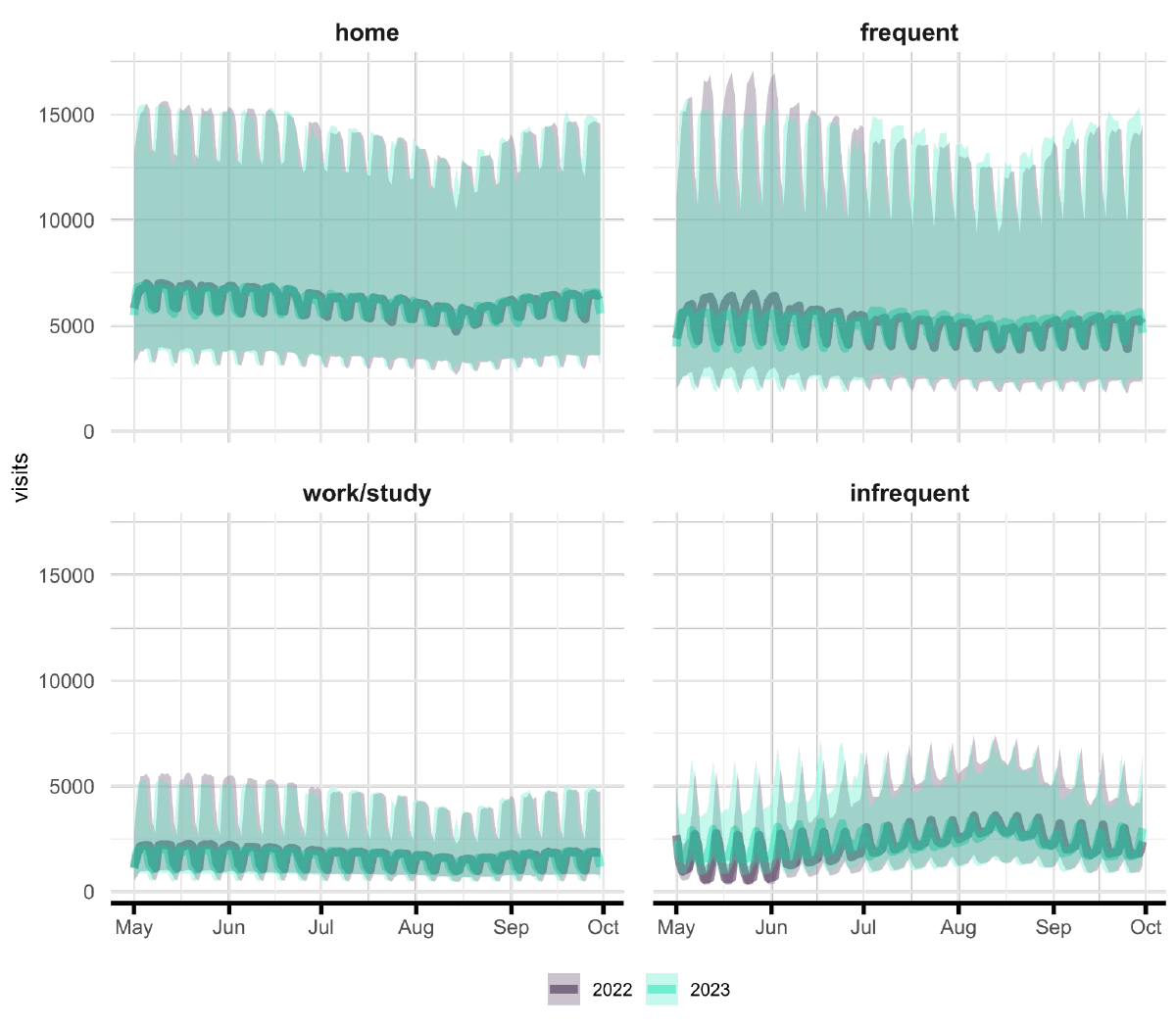}
\caption{\textbf{Trends in the data.} Time series per activity over the two summers of our study, with interquartile range shaded. We see that work/study and frequent activites are more common during the weak and infrequent activities are more common on the weekend; there is a seasonal trend wherein infrequent activities rise is August while work/study and frequent activities fall. We also note that in May 2022, when school is not in session, frequent activities are higher and infrequent activities lower than in the same period during 2023—possibly due to a classification error. In light of these trends, we make adjustments to our model specifications to account for these weekday/weekend and seasonal variations.}
\label{S1}
\end{figure*}

\begin{figure*}[h!]
\centering
\includegraphics[width=1\textwidth]{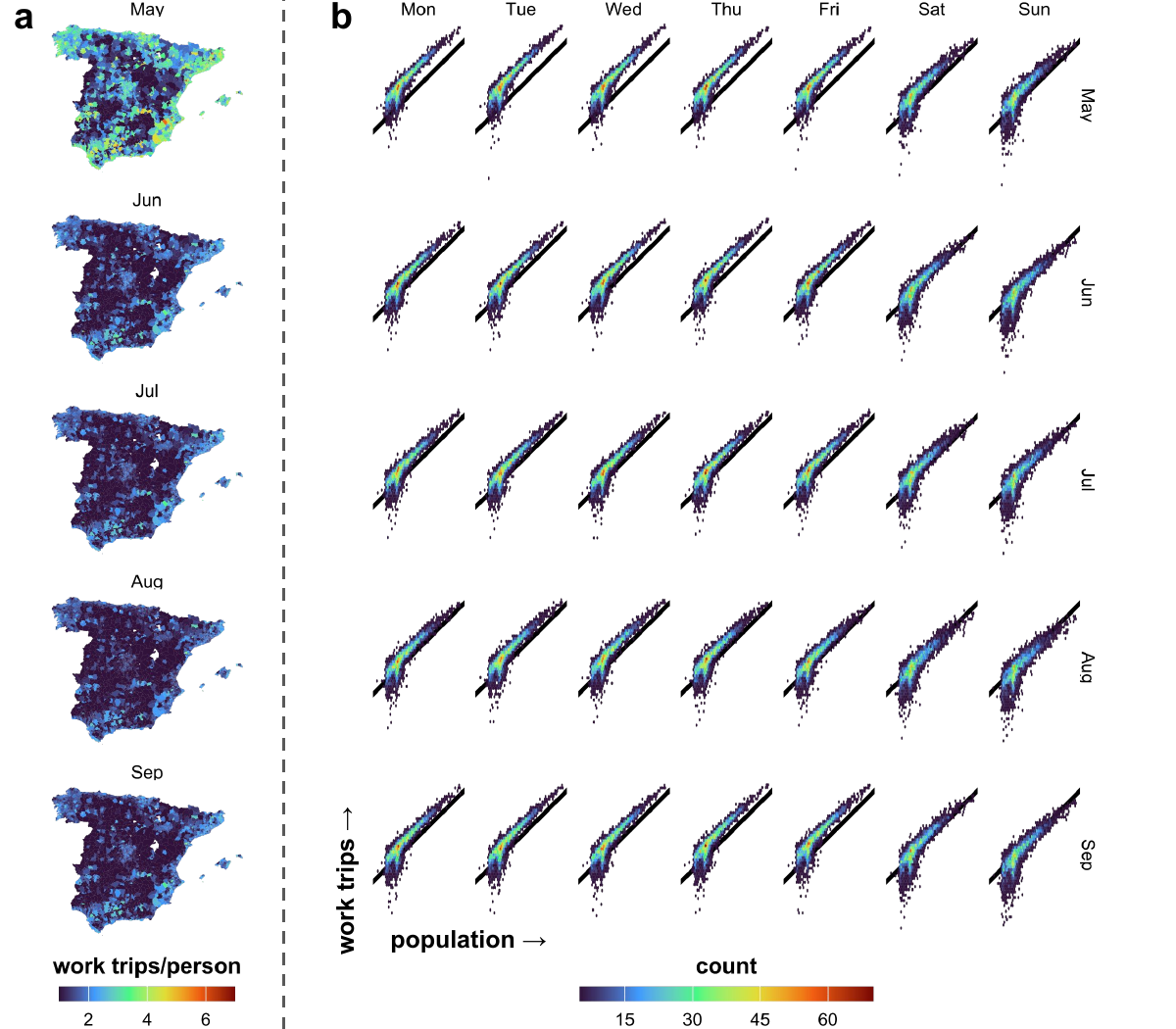}
\caption{\textbf{Mobility and population.} 
\textbf{A} The ratio work trips per person in each district for each month. We see that again in May the balance is different than in the remaining months; and urban districts tend to have more work trips than rural ones. This is because school is still in session in may, and work/study includes both of these trips. \textbf{B} We document a strong correlation between work trips and population, again with the exception of may, which typically has more work trips per person than other months. The fit returns to that of other months on the weekends, when school is out, during May. We address this in the model specification, and we also perform sensitivity analysis including and excluding this period.}
\label{S2}
\end{figure*}

\clearpage

\section{Model results}

\begin{table}[h!]
   \caption{Impact of Extreme Heat (UTCI $>40{^\circ}$C) on Mobility by Activity Type}
   \centering
   \normalsize
   \begin{tabular}{lcc}
      \midrule \midrule
      \textbf{Dependent Variable:} & \multicolumn{2}{c}{\emph{Flows}}\\
                                   & Within districts & Between districts\\
      \midrule
      \emph{Variables}\\
      Work or Study 
         & -0.0068                  
         & -0.0054                  \\
         & (0.0087)                 
         & (0.0079)                 \\
      Frequent Activity 
         & -0.0207\textsuperscript{**} 
         & -0.0243\textsuperscript{***}\\
         & (0.0086)                  
         & (0.0069)                  \\
      Infrequent Activity 
         & -0.125\textsuperscript{***} 
         & -0.105\textsuperscript{***}\\
         & (0.0311)                  
         & (0.0232)                  \\
      \midrule
      \emph{Fixed effects}\\
      District                   & Yes & Yes\\
      Date                       & Yes & Yes\\
      \midrule
      \emph{Fit statistics}\\
      Pseudo R\textsuperscript{2}  & 0.978 & 0.972\\
      \midrule \midrule
      \multicolumn{3}{l}{\emph{Notes: Standard errors (in parentheses) are clustered by province.}}\\
      \multicolumn{3}{l}{\emph{Significance levels: *** \(p<0.01\), ** \(p<0.05\), * \(p<0.10\).}}\\
   \end{tabular}
   \vspace{10pt}
   \caption*{\textbf{Note:} Results from Poisson regressions including district and date fixed effects. 
             Estimates represent semi-elasticities of flows (i.e., logs of expected counts). 
             Standard errors are robust to clustering at the province level.}
\label{T1}
\end{table}

\clearpage

\section{Sensitivity analysis}

\begin{figure*}[h!]
\centering
\includegraphics[width=1\textwidth]{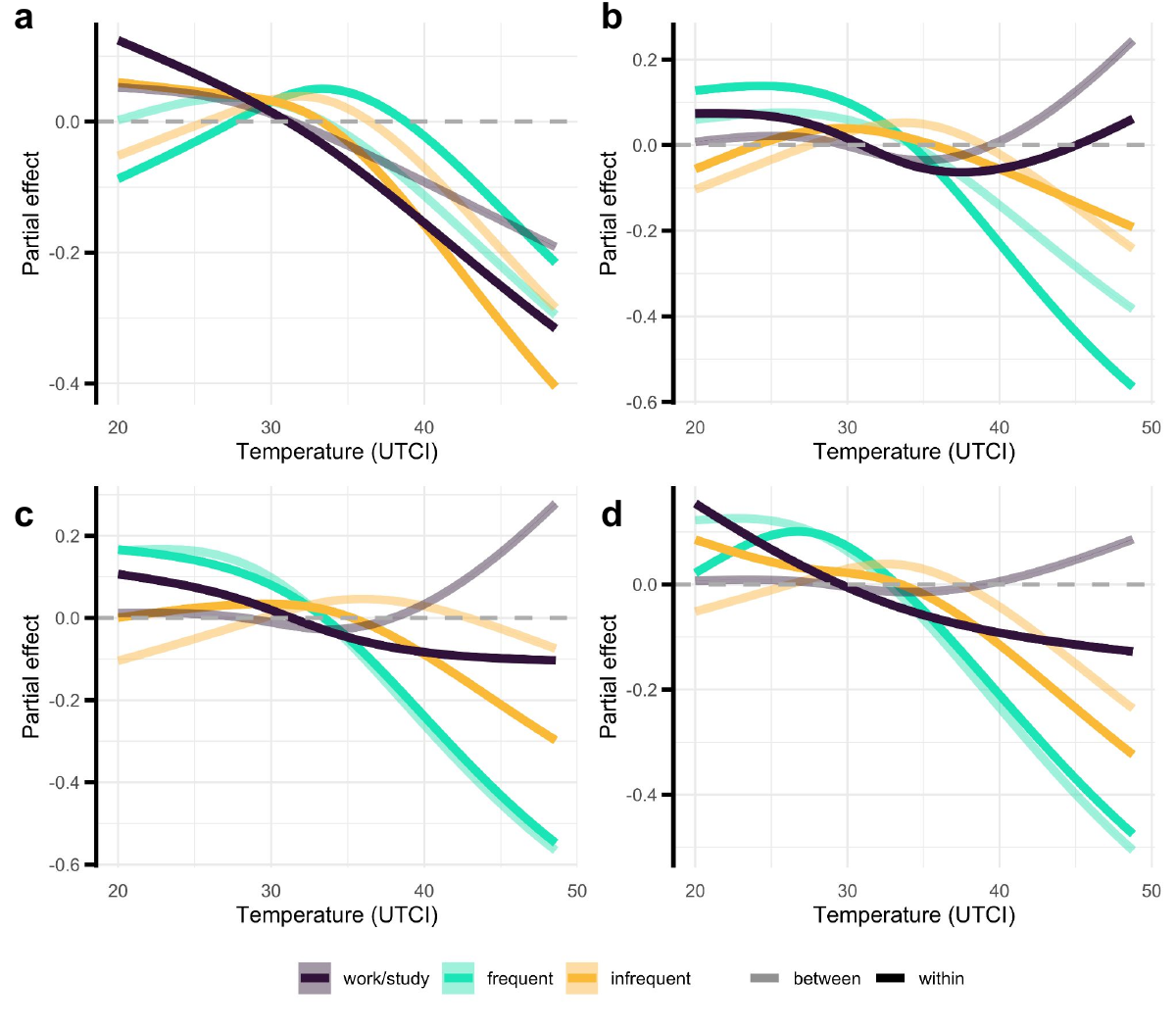}
\caption{\textbf{Sensitivity analysis.} 
\textbf{A} GAM results for the top 50 cities in Spain, all with a population over 100,000, and \textbf{B} for the remainder of Spanish cities, showing that the results are largely stable, with the possible exception being around work patterns: in large cities, people are more likely to avoid travel to work on hot days than in small cities—interesting because information technology jobs concentrate in large cities and thus can likely be taken from home. \textbf{C} and \textbf{D} filter the data to 2022 and 2023, respectively, showing that they are also stable across time.}
\label{S4}
\end{figure*}

\clearpage

\section{Placebo tests}
\begin{figure*}[h!]
\centering
\includegraphics[width=1\textwidth]{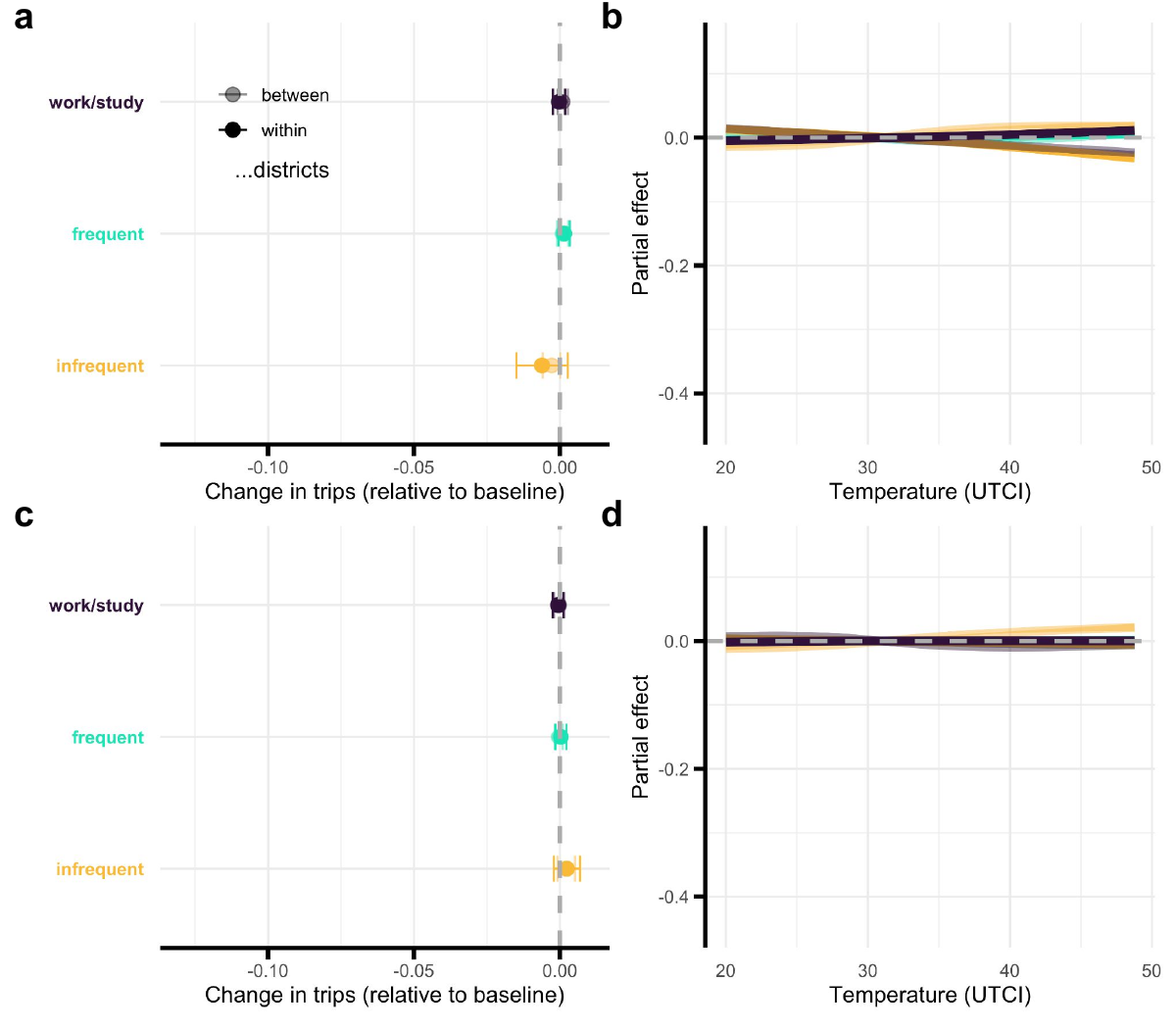}
\caption{\textbf{Placebo tests.} We perform 4 separate placebo tests to rule out potential spurious associations: \textbf{A} and \textbf{B} shuffle temperatures across districts on the same date, while \textbf{C} and \textbf{D} shuffle temperatures across different dates within the same district. The cross-district shuffles help rule out the possibility that our results are driven by events affecting all districts simultaneously. The within-district shuffles help rule out the possibility that observed changes would occur within districts regardless of temperature variation. While GAM confidence intervals occasionally exclude zero, effect sizes are reduced by an order of magnitude or more compared to the main analysis, supporting our primary findings.}
\label{S5}
\end{figure*}

\clearpage

\section{Consistency between genders}
\begin{figure*}[h!]
\centering
\includegraphics[width=1\textwidth]{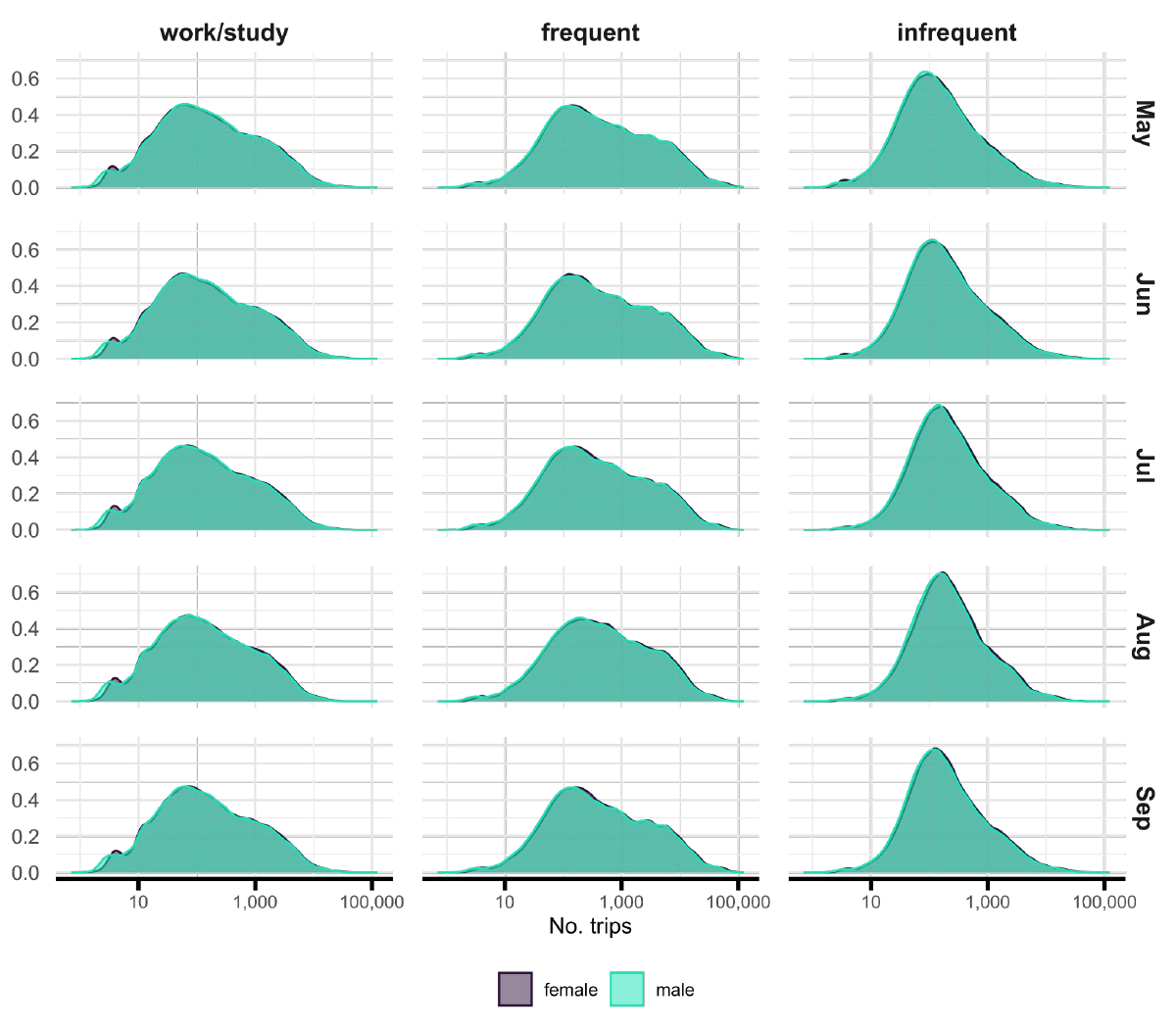}
\caption{\textbf{Trip distributions.} Here we show the distributions of daily visits per month and class of trip. We note that across all months and classes of trip, the distributions of trips are consistent between genders, suggesting broad similarity.}
\label{S6}
\end{figure*}

\clearpage

\section{Gravity model}

\begin{table}[h!]
   \centering
   \normalsize
   \begin{tabular}{lccc}
      \tabularnewline \midrule \midrule
      Dependent Variable: & \multicolumn{3}{c}{Trips}\\
                                          & Base model    & Temperature bins & With gradient \\   
      Model:                              & (1)           & (2)              & (3)\\  
      \midrule
      \emph{Variables}\\
      Log Distance                        & -1.87$^{***}$ & -1.87$^{***}$    & -1.87$^{***}$\\   
                                          & (0.013)       & (0.013)          & (0.013)\\   
      Temperature 25-30°C                 &               & 0.011$^{***}$    & 0.012$^{***}$\\   
                                          &               & (0.002)          & (0.002)\\   
      Temperature 30-35°C                 &               & 0.008$^{**}$     & 0.009$^{***}$\\   
                                          &               & (0.004)          & (0.004)\\   
      Temperature 35-40°C                 &               & -0.020$^{***}$   & -0.018$^{***}$\\   
                                          &               & (0.006)          & (0.006)\\   
      Temperature 40-45°C                 &               & -0.053$^{***}$   & -0.051$^{***}$\\   
                                          &               & (0.007)          & (0.007)\\   
      Temperature 45-50°C                 &               & -0.066$^{***}$   & -0.062$^{***}$\\   
                                          &               & (0.009)          & (0.010)\\   
      Log Distance × Temperature Gradient &               &                  & -0.001$^{*}$\\   
                                          &               &                  & (0.0007)\\   
      \midrule
      \emph{Fixed-effects}\\
      ID origin                          & Yes           & Yes              & Yes\\  
      ID destination                     & Yes           & Yes              & Yes\\  
      date                                & Yes           & Yes              & Yes\\  
      \midrule
      \emph{Fit statistics}\\
      Observations                        & 31,148,196    & 31,148,196       & 31,148,196\\  
      Pseudo R$^2$                        & 0.87382       & 0.87386          & 0.87386\\  
      Squared Correlation                 & 0.78625       & 0.78636          & 0.78637\\  
      \midrule \midrule
      \multicolumn{4}{l}{\emph{Clustered (origin \& destination) standard-errors in parentheses}}\\
      \multicolumn{4}{l}{\emph{Signif. Codes: ***: 0.01, **: 0.05, *: 0.1}}\\
   \end{tabular}
   
   \par \raggedright 
   \vspace{10pt} 
   \caption{\textbf{Impact of temperature on mobility.} We use gravity models to understand if flows are displaced, moving from one district to another, during heat waves, which would violate TWFE assumptions. While high temperatures reduce flows, agreeing with the TWFE analysis, we see that temperature gradient has no significant effect, suggesting that people do not preference higher or lower temperatures during heat waves. This also indicates that people do not seek out cooler parts of the country during extreme heat. All models include origin, destination, and date fixed effects; Standard errors clustered by origin and destination.}
\label{T2}
\end{table}

\begin{table}[h!]
   \centering
   \normalsize
   \begin{tabular}{lccccc}
      \tabularnewline \midrule \midrule
      & \multicolumn{5}{c}{\textbf{Population}}\\
      Temperature & Very Low & Low & Medium & High & Very High \\
      \midrule
      25-30°C & 0.0267$^{***}$ & 0.0324$^{***}$ & 0.0321$^{***}$ & 0.0376$^{***}$ & 0.0482$^{***}$ \\
              & (0.0031) & (0.0045) & (0.0086) & (0.0091) & (0.0092) \\
      30-35°C & 0.0589$^{***}$ & 0.0629$^{***}$ & 0.0371$^{***}$ & 0.0277$^{**}$ & 0.0224$^{**}$ \\
              & (0.0049) & (0.0059) & (0.0078) & (0.0080) & (0.0084) \\
      35-40°C & 0.0862$^{***}$ & 0.0834$^{***}$ & 0.0350$^{***}$ & -0.0007 & -0.0363$^{***}$ \\
              & (0.0075) & (0.0095) & (0.0077) & (0.0076) & (0.0077) \\
      40-45°C & 0.0791$^{***}$ & 0.0763$^{***}$ & 0.0086 & -0.0303$^{**}$ & -0.0874$^{***}$ \\
              & (0.0081) & (0.0108) & (0.0083) & (0.0089) & (0.0087) \\
      45-50°C & 0.0751$^{***}$ & 0.0588$^{***}$ & 0.0123 & -0.0783$^{***}$ & -0.0962$^{***}$ \\
              & (0.0158) & (0.0125) & (0.0130) & (0.0173) & (0.0146) \\
      \midrule
      log(d) & \multicolumn{5}{c}{-1.865$^{***}$ (0.0112)} \\
      \midrule
      \emph{Fixed-effects} & & & & & \\
      ID origin & \multicolumn{5}{c}{Yes} \\
      ID destination & \multicolumn{5}{c}{Yes} \\
      date & \multicolumn{5}{c}{Yes} \\
      \midrule
      \emph{Fit statistics} & & & & & \\
      Observations & \multicolumn{5}{c}{19,638,009} \\
      Pseudo R$^2$ & \multicolumn{5}{c}{0.87786} \\
      Squared Correlation & \multicolumn{5}{c}{0.79586} \\
      \midrule \midrule
      \multicolumn{6}{l}{\emph{Clustered (by origin \& destination) standard-errors in parentheses}} \\
      \multicolumn{6}{l}{\emph{Signif. Codes: ***: 0.001, **: 0.01, *: 0.05, .: 0.1}} \\
   \end{tabular}
   
   \par \raggedright 
   \vspace{10pt} 
   \caption{\textbf{Impact of temperature and combined population on mobility.} In this gravity model we report the results of a gravity model that groups flows by population at origin and destination. Because the model has 5x5 population classes as well as temperature bands, we present the average effects for combined population groups. Our results show large negative effects for the most populous areas, while moderate warmth is positive relative to cold.}
\label{T3}
\end{table}

\clearpage

\section{CMIP6 projections}
\label{projections}
\begin{figure*}[h!]
\centering
\includegraphics[width=1\textwidth]{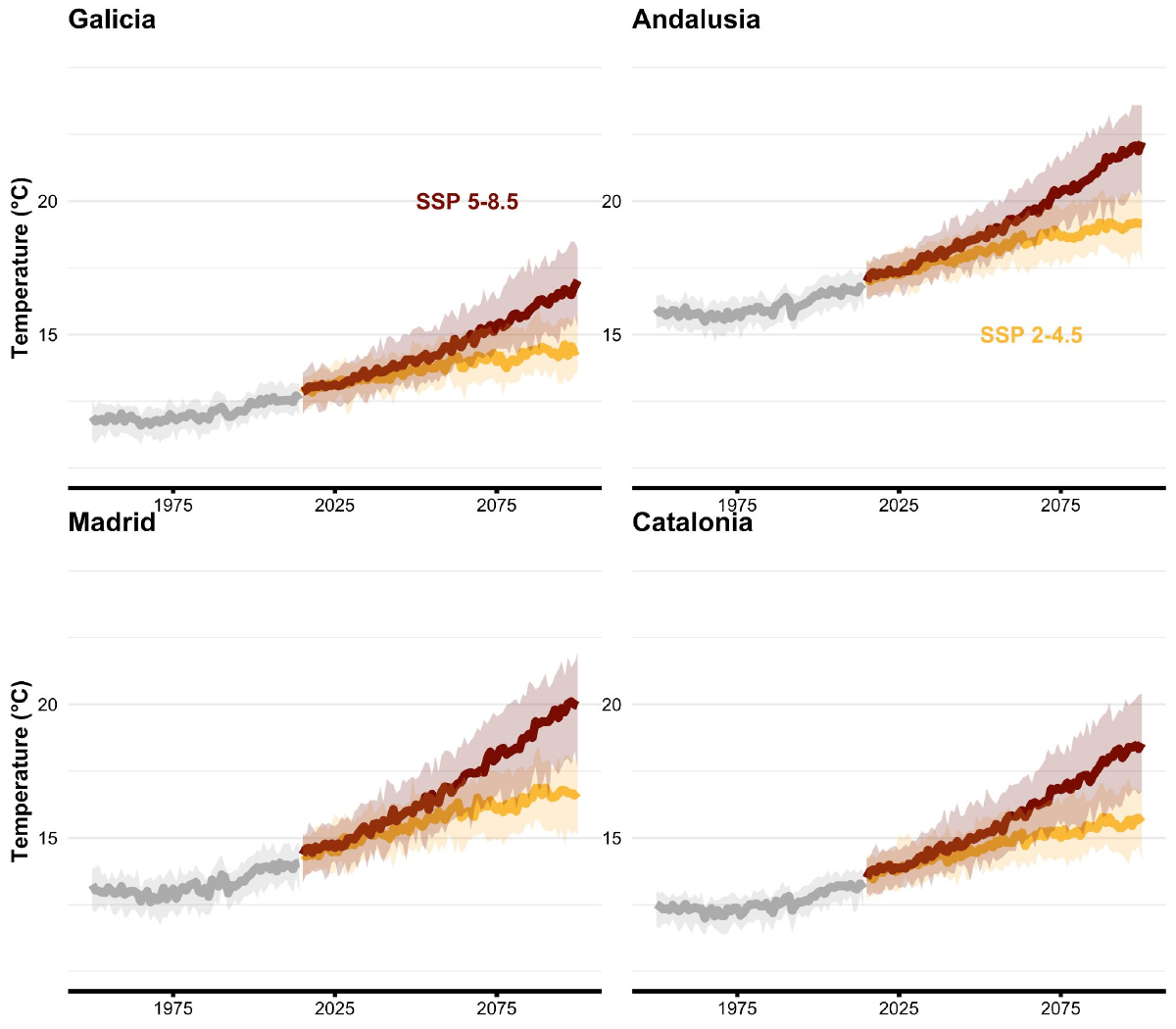}
\caption{\textbf{CMIP estimates.} Andalusia, in the South will see annual temperatures rise more than Galicia, in the North. Major cities like Barcelona and Madrid will also experience considerable changes.}
\label{S7}
\end{figure*}

\bibliographystyle{naturemag}
\bibliography{heat}